\documentclass[11pt,a4paper,twocolumn]{article}

\usepackage[a4paper,left=16mm,right=16mm,bottom=25mm,top=25mm]{geometry}
\usepackage{fancyhdr}
\usepackage{subfigure}
\usepackage{graphicx}
\usepackage{booktabs}
\usepackage[british]{babel}
\usepackage[square,comma,numbers,sort&compress]{natbib}
\usepackage{csvsimple}
\usepackage{graphicx}
\usepackage{pgfplotstable,filecontents}
\usepackage{gensymb}
\usepackage{array}
\usepackage{tabu}
\usepackage{multirow}
\usepackage{url}

\usepackage{xcolor}
\usepackage{afterpage}
\usepackage{graphicx}
\usepackage{dcolumn}
\usepackage{bm}
\usepackage[title]{appendix}
\usepackage[]{subfigure}
\usepackage{float}
\usepackage{hyperref}
\hypersetup{colorlinks,linkcolor={blue},citecolor={blue},urlcolor={cyan}}
\usepackage{amsmath}
\usepackage{cleveref}
\usepackage{afterpage}

\begin{document}

\title{\Large Tailoring electronic properties on Bi$_2$O$_2$Se under surface modification and magnetic doping}
\author{I. Arias-Camacho$^{1,2}$, A.M León$^{1,3}$, J. Mejía-López$^1$}

\date{%
$^1$Facultad de Física, Pontificia Universidad Católica de Chile, Santiago, Chile. \\%
$^2$ Institute of Experimental Physics, University of Warsaw.\\%
$^3$Max Planck Institute for Chemical Physics of Solids, Dresden, Germany.\\[2ex]%
\today
}

\twocolumn[
\begin{@twocolumnfalse}
\maketitle
\begin{abstract}
The search for a two-dimensional material that simultaneously fulfills some properties for its use in spintronics and optoelectronics, i.e., a suitable bandgap with high in-plane carrier mobility and good environmental stability, is the focus of intense current research. If magnetism is also present, its range of utility is considerably expanded. One of the promising materials fulfilling these features is Bi$_2$O$_2$Se, a non van-der-Waals system whose monolayer has been recently obtained. This study addresses the structure and electronic properties of different monolayers that could be obtained experimentally. It is observed that these monolayers are very sensitive to the introduction of ``extra" electrons, changing their electronic character from semiconductor to conductor. Furthermore, we investigate how the properties of each studied monolayer change when the system is doped with magnetic atoms. The result is that the doping introduces bands of low dispersion caused by the d orbitals of the impurities that can hybridize with the oxygen and bismuth atoms in the monolayer. This strongly modifies the electronic properties of the material, producing changes in the valence of certain Bi atoms which can induce a symmetry breaking in the perpendicular plane. Such phenomena lead to metallic or semiconducting characteristics, depending of metal doping.\\
\end{abstract}
\end{@twocolumnfalse}]

\section{\label{sec:level1}Introduction}

2D van der Waals (vdW) materials provide a large variety of correlated electronic-optical and magnetic properties, enabling the design of novel multifunctional heterostructures  \cite{Wang20} \cite{Jiang21}. Experimentally, these materials are obtained due to the weak interaction (vdW) existing among the different layers that compose their corresponding bulk. Beyond the vdW materials, a new family of 2D systems (obtained from Sillen structures \cite{Sillen}) has been recently investigated, which are characterized by a strong ionic interlayer bonding, and, therefore, they are not vdW materials. The interest in these materials has increased rapidly and, to date, it has been predicted the existence of around 28 exfoliated compounds \cite{ricco}. Unlike vdW materials, they present dangling bonds and unsaturated coordination sites, enabling a high degree of surface chemical activity \cite{ricco,jin2021two}, which offers excellent capabilities for carrier transfer, sensing, catalysis, and even magnetism. However, their synthesis, as well as achieving their stability in air condition is still a challenge.

The bismuth oxychalcogenides systems (Bi$_{2}$O$_{2}$X, X $=$ S, Se and Te) belong to the family of layered non-vdW materials, and they are considered fascinating materials due to their unique electronic properties \cite{Optoelectronic2017} . They are excellent semiconductors with a bandgap between 0.4-1.3 eV, depending on X and the layer number. Besides, they are characterized by presenting a low intrinsic in-plane thermal conductivity and high electrical conductivity \cite{Ruleova2010}\cite{Guo2013}\cite{Wu2017}\cite{Synthesis17}. It is believed that the combination of such properties is due to the strong interlayer coupling, which leads to a strong anharmonicity of phonons, lowering the thermal conductivity \cite{wang2018electron}. Among these materials, Bi$_{2}$O$_{2}$Se (BOS) has been of particular interest given its ultrahigh Hall mobility, whose value varies from bulk to the bi-layered (BL) system. In fact, it has been observed that BOS nanoflakes exhibit Hall mobility  $>$ 20.000 cm$^{2}$ V$^{-1}$ s$^{-1}$ at room temperatures, being this value comparable with the observed in graphene, and below the bilayer limit of around 450 cm$^{2}$ V$^{-1}$ \cite{wu2017high}. Regarding its thermal properties, this material presents high Seebeck coefficients, varying between 150 and 240 $\mu$V K$^{-1}$ at T = 300 K; and its merit factor (ZT) value is 3.35 at 800 K \cite{yu2018bi2o2se}, larger than that observed in SnSe (ZT $=$ 2.6), which is known as the most efficient thermoelectric material \cite{zhao2014ultralow}. All these properties make BOS systems excellent candidates for optical-thermoelectric devices. Additionally, recent experiments reported that BOS shows a ferroelectric transition at 235 $^{\circ}$C, and that its polarization can be tuned by applying an electric field. This last discovery opens new perspectives for maximizing the thermoelectric performance of BOS by changing the scattering mechanism and carrier mobility over a wide temperature range \cite{ghosh2019ultrathin}. 

Although BOS thin films have been successfully grown via CVD in recent decades \cite{Wu2017, RobustBG18, Strain2019}, there is some controversy in finding the most suitable structure to model its isolated monolayer. The problem lies in the unsaturated bonds that appear on the surfaces due to the ionic character of the system, which makes the structure unstable. One model used to keep the stability while preserving the bulk geometry consists in the passivation of the Se atoms (which are located on both monolayers). This is achieved by using hydrogen atoms to avoid those unsaturated bonds \cite{Strain2019,Legut2019}. Nevertheless, it turns out to be invalid in this specific case since it does not reproduce adequately the gap dependence while increasing the number of monolayers, and, although it would preserve the system stability, it is known that there is no hydrogen involved in the experiments \cite{Wei2019}.
 
Another model of the monolayer was proposed by Wei et al. \cite{ Wei2019}. Their model conserves the stoichiometry of the bulk by removing two atoms from the upper and bottom planes (Zipper model). It was based on the observation of Chen et al. \cite{RobustBG18}, who cleaved BOS monocrystals inside ultrahigh vacuum chambers (cleavage occurs in the Se plane) and observed by STM measurements that 50\% of Se atoms remained attached to each BOS plane. In this case, Se atoms were arranged in dimmers alternated with one vacancy (V) in the sequence V-Se-Se-V. This model correctly reproduces the thickness dependence of the band-gap.

In these experimental and theoretical studies, it is observed that the BOS does not possess intrinsic magnetism. However, recent studies based on DFT predicted the emergence of long-range magnetic order at near room temperature in BOS doped with 3d transition metals such as Mn, Fe, Co and Ni \cite{Legut2019}. Due to its excellent air stability and their unique electronic properties, the inclusion of magnetism could offer new perspectives toward their use in spintronics and could also open a new stage in the exploration of BOS-based magnetic compounds.

Motivated by these recent advances and the controversies on the nature of the BOS surface, in this work, we investigate different arrangements of monolayers that are stable and that could be candidates for the fundamental structure of a BOS monolayer. We study their electronic properties and the effects of introducing magnetic impurities (atoms of Mn, Fe, Co and Ni).

\section{Model and Methodology}

First-Principles calculations were performed by using projector augmented wave (PAW) pseudopotentials \cite{PAWmethod94,PAW2nd} as implemented in the \textit{Vienna Ab initio Simulation Package} (VASP) \cite{VASP96,VASP2nd}. 
For the study of the monolayers, a 25 Å vacuum layer (along the z-direction) was considered in order to avoid  interactions between adjacent layers.
The kinetic energy cutoff for plane waves was set at 500 eV, while the k-point grid in the Brillouin zone, within the Monkhorst-Pack scheme, was set at 11x11x11 for the bulk and 9x9x1 for the monolayer. These values ensure the accuracy of the total energy calculations. 
The K-mesh was increased to 17x17x1 to perform density of states (DOS) calculations; and full relaxation was performed until the residual forces on the atoms were below 0.008 eV/Å. The total energy convergence criterion was set to 10$^{-8}$ eV/Å. 
Correlation and electron exchange effects were treated with the Generalized Gradient parameterized Perdew-Burke-Ernzerhof (PBE) \cite{PBE96} approximation. The d-states of transition atoms were treated with the LDA+U method introduced by Dudarev \cite{Dudarev98} to correct the repulsive interaction by introducing the U $-$ J (= 5 eV) parameter to increase the repulsive Coulomb potential, as it was considered in previous works \cite{Legut2019}.
 All the visualizations were performed in the Visualization for Electronic and structural Analysis software (VESTA) \cite{VESTA}.
 
For the structural characterization of the systems, we use the cohesive energy per atom, which is the difference in energy between the formation energy of the compound and the sum of the formation energies of the isolated atoms; thus:
 
\begin{multline}
E_{cohesive} = (E[Bi_{2}O_{2}Se] - n_{Bi}E[Bi] - n_{Se}E[Se] - \\
n_{O}E[O])/(n_{Bi}+n_{Se}+n_{O}) \\
\end{multline}

where E[Bi], E[O] and E[Se] are the total energies of Bi, O and Se, respectively; $n$ is the number of atoms per unit cell; and E[Bi$_{2}$O$_{2}$Se] is the total energy of the monolayer.
With this definition, negative cohesive energy implies that the energy to create isolated atoms is greater than that needed to form the complex, and, therefore, the more negative the cohesive energy of a system, the more energetically stable it is. Since the number and nature of the species involved remain unaltered, this quantity does not depend on the chemical potentials.


The doping consists in replacing the central Bi atom with a transition metal (TM) atom of each impurity, as shown in Fig. \ref{fig:Doping}. Since our supercell has eight Bi atoms, then we have 12.5\% as our doping percentage.

\begin{figure*}[ht]
\centering
\includegraphics[width=19cm]{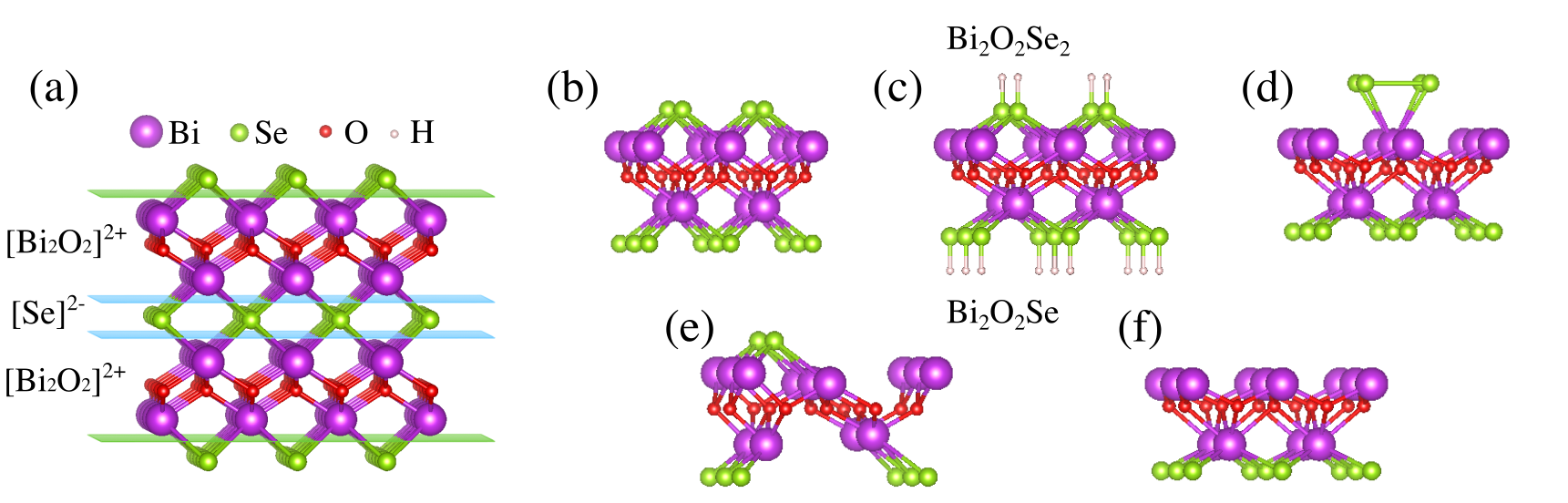}
\caption{Relaxed structures for bulk and the monolayer models studied. Bulk Bi$_2$O$_2$Se (a). The non-stoichiometric systems: P-terminated (b), Se-terminated (c) and R-Se-terminated (d). The stoichiometric systems: Zipper (e), and One-Se-plane (f).}
\label{fig:Structures}
\end{figure*}

The qualitative comparison of the electrical conductivity between different metallic systems is based on the Drude's model \cite{Hummel}:

\begin{equation}
 \sigma = \frac{ne^2\tau}{m} = \frac{1}{3} e^2 v_F^2 N(E_F)\tau
\end{equation}

\noindent where $n$ is the electronic density, $e$ and $m$ the electron charge and mass, respectively; $\tau$ is the relaxation time, $v_F$ the Fermi velocity, and $N(E_F)$ is the DOS at the Fermi energy $E_F$. In other words, the conductivity is proportional to the DOS occupied at the Fermi level.

\section{Results and Discussion}

\subsection{Pristine Monolayers}

\subsubsection{Structural Properties}

The structure of bulk Bi$_2$O$_2$Se is a tetragonal lattice that belongs to the I4/mmm space group (No. 199) and it presents a layered nature. 
In this particular compound, each oxygen atom bonds covalently with the four surrounding Bi atoms to form a Bi$_4$O tetrahedron, and such tetrahedra arrange themselves to create the positively charged [Bi$_2$O$_2$]$^{2+}$ layers, which alternate with the negatively charged [Se]$^{2-}$ layers (Fig. \ref{fig:Structures}a). 
The resulting lattice parameters for bulk after optimization are 
$a$ = $b$ = 3.92 Å, $c$ = 12.38 Å, which are in very good agreement with experimental data ($a$ = $b$ = 3.89 Å, $c$ = 12.21 \cite{Drasar2012} Å and $a$ = $b$ = 3.88 Å, $c$ = 12.16 Å \cite{Synthesis17}).

The different models of the monolayers used in our study were built according to the aforementioned bulk geometry, and they are shown in Figs. \ref{fig:Structures}b-f. We have divided them into two groups called stoichiometric (Bi$_2$O$_2$Se) and non-stoichiometric (Bi$_2$O$_2$Se$_2$) regarding their bulk counterpart. The structural parameters obtained from the DFT calculations for these monolayers are shown in Table \ref{tab:bond}

The first two non-stoichiometric structures studied have been named Se-terminated (Fig.\ref{fig:Structures}b) and P-Se-terminated (Fig. \ref{fig:Structures}c). They were built to preserve the bulk symmetry, and it is worth recalling that they have been the most common models used in several theoretical works \cite{Legut2019,Zhu2019,Strain2019,MecFlexibility20,Hu2021}. 
For this ML, we used a 2x2x1 supercell containing 24 atoms, 8 of each species (see Fig. \ref{fig:Structures}b), with atomic planes equidistant to the plane of oxygen. After a full relaxation, the resulting lattice parameters $a$ and $b$ are 4\% smaller (3.77 Å) than for the bulk case (3.92 Å).
This effect causes a slight shortening of the Bi-Se and Bi-O bonds with respect to the bulk, as shown in Table \ref{tab:bond}.
To stabilize this structure so that it resembles one of the bulk monolayers, some works used hydrogen atoms to passivate the unsaturated bonds of the Se atoms \cite{Synthesis17,Thermo18,RobustBG18}. We have also performed the optimization of this passivated monolayer (P-Se-terminated, Fig. \ref{fig:Structures}c) for comparison.
In this case, the obtained lattice parameter is 4.02 Å; 2.5\% larger than the bulk, although its Bi-Se bonds are shorter (3.8\%) and Bi-O bonds have increased 1.5\% respect to the bulk system. We can see that the passivated monolayer has a higher cohesive energy ($\approx$ 16\%, see Table \ref{tab:bond}) than the Se-terminated ML monolayer. Thus, it can be said that the latter one is a more stable structure. 

\begin{table*}[ht!]
\begin{center}
\resizebox{17.7cm}{!} {
\begin{tabular}{| c | c | c | c | c | c | c | c | c |}
\hline
Structure & $a$ (\AA) & $b$ (\AA) & $c$ (\AA) & $d_{Bi-Se}$ & $d_{Bi-O}$ &Cohesive & $\nu$ at $\Gamma$ (cm$^{-1}$) & el. effective \\
 &  &  &  & (\AA) & (\AA) & energy (eV) & & mass (m$_{e}$) \\
\hline \hline
Bulk &  3.92  & 3.92 & 12.38 & 3.3 & 2.3 & -2.561 & 489.9 & -0.509 (0.073) \\ \hline \hline
Se-terminated ML & 3.77 & 3.77 & 6.08* & 3.1 & 2.3 & -2.270 & 465.7 & - \\ 
P-Se-terminated ML & 4.03 & 4.03 & 8.40* & 3.2 & 2.4 & -1.914 & 2347.2 & -1.357 (0.153)\\ 
R-Se-terminated ML & 3.88 & 3.92 & 6.73* & 3.1/3.3 & 2.5/2.2 & -2.426 & 591.1 & -\\ 
Zipper ML & 3.89 & 4.21 & 5.80* & 2.8/3.4 & 2.5/2.3 & -2.454 & 521.8 & -0.882 (0.275)\\ 
One-Se-plane ML & 3.96  & 3.96 & 3.78* & 3.0 & 2.2/2.6 & -2.376 & 588.3 & -0.178 (0.313) \\ \hline
\end{tabular}}\\
(*)Thickness of the ML
\caption{Lattice parameters, bond lengths for Bi-Se and Bi-O, cohesive energies, the highest frequencies at the $\Gamma$ point and electron effective mass (m$_{e}$) in B.V. (B.C.), for the Bi$_2$O$_2$Se bulk and the different monolayer models.}
\label{tab:bond}
\end{center}
\end{table*}

The other non-stoichiometric monolayer investigated in this work is a reconstruction of the Se-terminated ML (R-Se-terminated, Fig. \ref{fig:Structures}d), which resulted after doping it with the transition metal (TM) and replacing again the dopant with the Bi atom; as will be described in detail in section B. This R-Se-terminated ML has a lower cohesive energy with respect to the first one (156 meV lower) accompanied by a rearrangement of the upper Se atoms in a tetramer and the approach between the bottom planes of Bi and Se in 28\% with respect to Se-terminated ML.

One of the stoichiometric Bi$_2$O$_2$Se ML considered here has the shape of a Zipper as proposed by Wei et al.\cite{Wei2019}, and consists of removing four Se atoms from the Se-terminated ML: two atoms from the upper plane and two atoms from the bottom plane; as shown in (Fig. \ref{fig:Structures}e). 
The lattice parameters obtained from our calculation result to be asymmetric, $a$ = 3.934 Å and $b$ = 4.086 Å, that is, they are expanded a 0.3\% and 4.2\% with respect to the bulk. This monolayer has the larger cohesive energy among those we report here ($\approx$ 8\% lower than Se-terminated).

Another different stoichiometric ML is shown in Fig. \ref{fig:Structures}f (which will be called One-Se-plane hereafter), it is the result of removing one of the Se planes from the Se-terminated ML.
In this case, although Bi-Se and Bi-O are the shortest bonds among all the 
MLs studied here (with reductions of 9\% and 4.3\% as compared to the bulk case), the lattice parameters are in close proximity with those of the bulk, with a slight increase of just 1\%. 
Besides, for  this One-Se-plane ML, cohesive energy is 4.5\% smaller than that of the Se-terminated ML and 3.2\% larger than in the Zipper case.

It is worth noting that the R-Se-terminated and the Zipper monolayers do not preserve the same value for the in-plane lattice parameters, $a$ and $b$, being different by $\approx$1\% and $\approx$8\%, respectively (see Table \ref{tab:bond}). We have also studied the bilayer (BL) of the Zipper ML proposed by Wei et al. (Fig. \ref{fig:ZipperBL}), obtaining as a result that the addition of another monolayer brings back the in-plane structure of the bulk phase. This is due to the completion of the selenium row in the center, which avoids the distortions suffered by the Zipper ML.\\

\begin{figure}[h]
\centering
\includegraphics[width=0.45\textwidth]{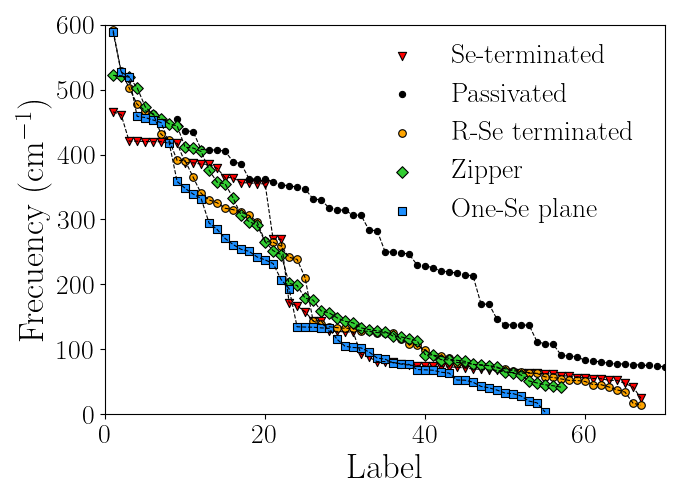}
\caption{Normal modes at the $\Gamma$ point for the four models of MLs.}
 \label{fig:VibrationalModes}
\end{figure}

To verify the stability of the MLs proposed, we have computed the frequencies at the $\Gamma$ point, confirming that all of these models are stable states, with no imaginary frequencies in each case. Fig. \ref{fig:VibrationalModes} compares the normal vibrational modes at the $\Gamma$ point of each structure, and the values of the highest frequency for each monolayer are also shown in Table \ref{tab:bond}. We can observe that the frequencies for the R-Se-terminated (591.15 cm$^{-1}$), Zipper (521.78 cm$^{-1}$) and One-Se-plane (588.32 cm$ ^{-1}$) monolayers are higher than that of the bulk (489.93 cm$^{-1}$), which reinforces the idea that a surface reconstruction with stronger bonds at the surface would have occurred to these particular models. The high-frequency value in the passivated system (2347.2 cm$^{-1}$) is due to a vibrational mode involving H atoms. The value of the highest frequency in which the H atoms do not participate is 454.94 cm$^{-1}$.

Results reported in Table \ref{tab:bond} show that the Zipper ML would be the most stable (-2.454 eV). In addition, since all the structures under study are stable, they could be obtained experimentally, depending on the general synthesis conditions and, mostly, on the 
substrate.

\subsubsection{DOS and Band Structure}

\begin{figure}[h]
\centering
\includegraphics[width=0.45\textwidth]{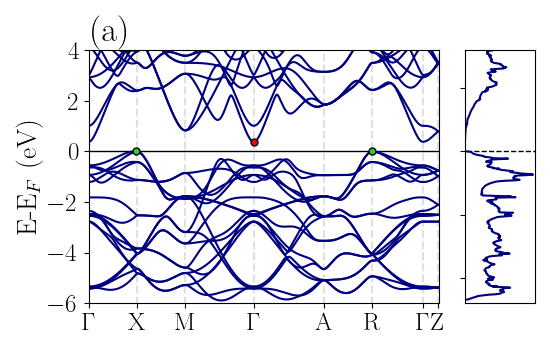}
\includegraphics[width=0.45\textwidth]{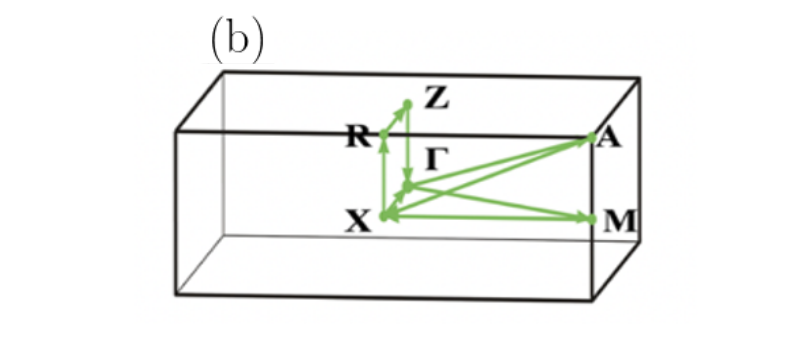}
\caption{(a) Band structure and DOS of the bulk system. The red and green points show the minimum of the conduction band (CBM) and the maximum of the valence band (VBM), respectively, indicating a semiconducting behavior. (b) Highly symmetric K-points, used in the calculation of the band structure, schematized in the first Brillouin zone of the unit cell.}
\label{fig:BandsBulk}
\end{figure}

Figure \ref{fig:BandsBulk} shows the band structure and DOS of Bi$_2$O$_2$Se in its bulk phase, which has been calculated for comparison purposes. 
We observe that it corresponds to an indirect bandgap semiconductor, with the maximum of the valence band located at the X point (green point in Fig.\ref{fig:BandsBulk}) and the minimum of the conduction band at the $\Gamma$ point (red point in Fig.\ref{fig:BandsBulk}). These results are in good agreement with the experimental values previously reported \cite{Wu2017,RobustBG18}. However, in our calculations we obtain a smaller value of the gap, 0.49 eV, with respect to the experimental value of 0.8 eV. This is due to the well-known gap underestimation of the PBE method \cite{Gap_underestimation}.
The valence band of the bulk system starts at -6.0 eV and it is dominated by the contributions of Bi and O atoms up to -2 eV. From there on, there is a contribution dominated by the p orbitals of the Se atoms, which begins to predominate. From the decomposition of the band structure into orbitals of each atom (not shown), we observe that the valence band maximum (VBM) is dominated by p$_{x}$ and p$_{y}$ orbitals of the Se atoms, whereas the minimum of the conduction band (CBM) corresponds to the p$_{z}$ orbitals of the bismuth.
We also note that there is a greater band dispersion in the plane of the layers ($\Gamma$-X-M-$\Gamma$) than out-of-plane ($\Gamma$-Z and A-R), meaning that the in-plane transport properties are more important.

\begin{figure*}[ht]
\includegraphics[width=0.95\textwidth]{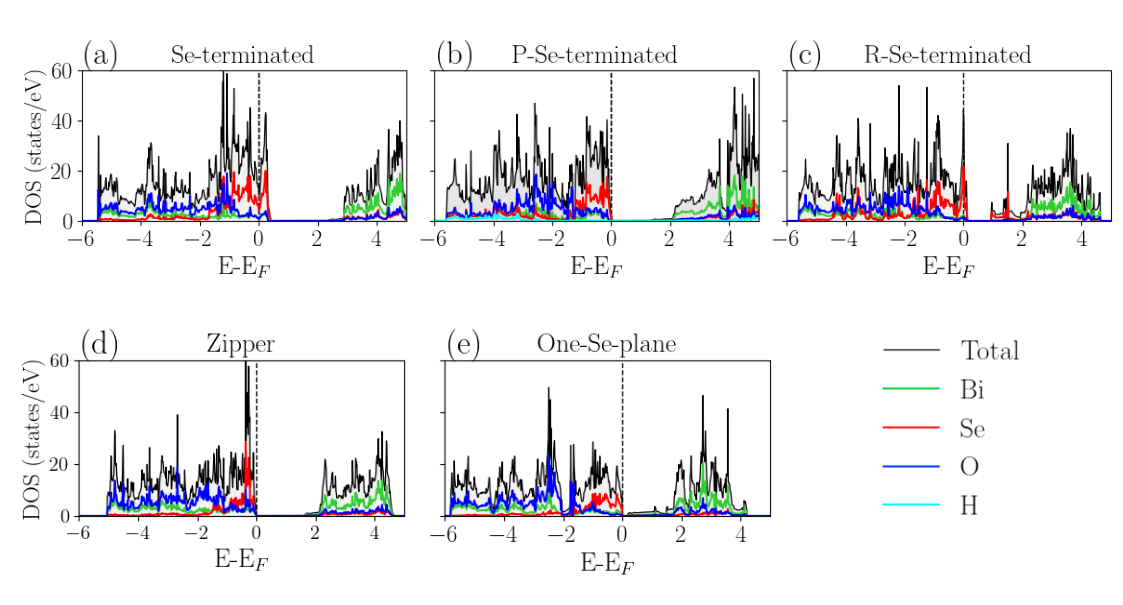}
\caption{DOS of each monolayer: The upper panel shows the non-stoichiometric systems: (a)-(b) Se-terminated and P-Se-terminated, respectively, and (c) R-Se-terminated. The lower panel shows the stoichiometric systems: (d) Zipper and (e) One-Se-plane.}
\label{fig_dos-mono}
\end{figure*}

Figure \ref{fig_dos-mono} shows the total and partial DOS of each ML. It is observed that in the cases of Se-terminated (Fig.\ref{fig_dos-mono}a) and R-Se-terminated (Fig.\ref{fig_dos-mono}c) monolayers, the change from bulk to monolayer resulted in important changes in the electronic structure of the material, i.e., from semiconducting to 
metallic behavior. On the other hand, the passivated monolayer (Fig.\ref{fig_dos-mono}b) as well as the stoichiometric Zipper and the One-Se-plane monolayers (Fig.\ref{fig_dos-mono}d and Fig.\ref{fig_dos-mono} e) remain as semiconductors with bandgaps of 1.31 eV, 1.61 eV and 0.16 eV, respectively. Thus, qualitatively, we found that the corresponding bandgaps of the P-Se-terminated and Zipper MLs increase 2.7 and 3.3 times, respectively, as compared to the bulk case. Conversely, the bandgap for the One-Se-plane ML decreases approximately to a third of the bulk value. Therefore, taking into account that the experimental bandgap of the bulk is 0.8 eV, these proportions would lead to experimental bandgap values of around 2.16, 2.64, and 0.26 eV, respectively. 

The most important contributions of the last occupied levels, near E$_{F}$ for the metals and at the edge of the valence band for the semiconductors, come from the p$_{x}$ and p$_{y}$ orbitals of the Se atom (Fig.\ref{fig_dos-mono}a-d), while the p$_{z}$ orbitals from the Bi atom dominate the unoccupied levels.
The non-stoichiometric MLs have a greater contribution of the Se atoms, then a greater number of electrons occupy the orbitals of the system. Such increase produces a shift of the density of states of Se towards positive energies, and a higher hybridization between Bi and O atoms. It would explain the metallic behavior of Se-terminated and R-Se-terminated MLs, which, from other works, appears to be very sensitive to small changes in carrier concentrations or under load \cite{Metallicity2021,Ferroelectricity2019,MechandStrain21,Tong2018}.
In the case of the passivated system, which also has an excess of Se atoms, the additional hydrogen atoms prevent a charge transfer from the Bi atoms to the Se atoms, as confirmed by the Bader's analysis of the charge distribution (Table \ref{tab:Bader1}), which shows the transfer of an electron from the Se to the H atoms. This mechanism reduces the "extra" contribution from excess Se atoms and this monolayer remains a semiconductor just like the bulk. 
Also, since the Zipper and the One-Se-plane MLs do not have excess Se atoms, they remain in a semiconducting 
state.

\begin{figure*}[ht!]
\centering
\includegraphics[width=0.95\textwidth]{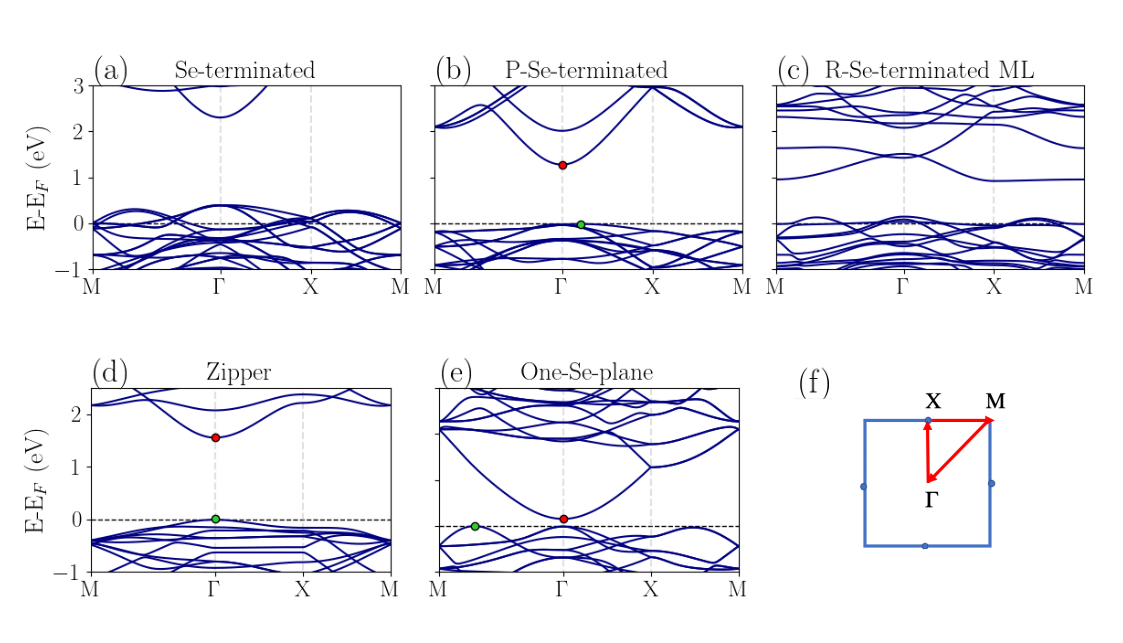}
\caption{Band structure for the following systems: (a) Se-terminated ML, (b) P-Se-terminated ML, (c) R-Se-terminated ML, (d) Zipper ML and (e) One-Se plane ML. (f) Highly symmetrical points shown in the first Brillouin zone.}
\label{fig:BandsUndopped}
\end{figure*}

Figure \ref{fig:BandsUndopped} shows the band structure for the monolayers investigated.
As observed, the passivated ML exhibits a semiconductor behavior, similar to the bulk case, 
with an indirect gap but with the VBM located in between the $\Gamma$ and X points, 
instead of at X. Moreover, the nature of the orbitals involved in the electronic structure does 
also coincide with that in the bulk structure.
In addition, the electron effective mass in the VBM of the passivated monolayer (-1.357 m$_e$, see Table \ref{tab:bond}) is much larger than the bulk one (-0.509 m$_e$), which indicates a drop of the electron mobility. 

In the case of the stoichiometric monolayers, the Zipper is the one that shows the greatest discrepancies with respect to the bulk since in the valence band, around the $\Gamma$ point, a hybridization of the p$_x$ and p$_y$ orbitals of selenium and oxygen takes place. Although this ML preserves the semiconducting nature of the bulk material, its bandgap is direct (instead of the indirect gap of the bulk, Fig. \ref{fig:BandsBulk}(a)) with both VBM and CBM located at the $\Gamma$ point.  In the case of the bilayer (see Figure \ref{fig:ZipperBL}), the direct bandgap diminishes from 1.60 eV to 0.62, what reproduces adequately its thickness dependence, which has also been reported in other theoretical \cite{Wei2019} and experimental \cite{MBE19,Wu2017} works. However,  in the latter, ARPES measurements show a thickness-dependent indirect bandgap between the $\Gamma$ point and at around the X point, which arises from delocalized out-of-plane orbitals p$_z$ and localized in-plane p$_x$ and p$_y$ orbitals.

The One-Se-plane ML shows the same trends as the bulk but with a smaller indirect bandgap. If the respective electron effective masses are compared, the One-Se-plane ML shows a smaller value (-0.178 m$_e$) than the Zipper ML (-0.882 m$_e$), indicating higher electron mobility for the former. Such values are found above and below the corresponding bulk value (-0.509 m$_e$).

\begin{table*}[ht]
\begin{center}
\begin{tabular}{|c|c|c|c|c|c|}
\hline
\multicolumn{1}{ |c| }{Model of monolayer} & \multicolumn{1}{ |c| }{Parameter} & \multicolumn{1}{ |c| }{Mn-doped} & \multicolumn{1}{ |c| }{Fe-doped} & \multicolumn{1}{ |c| }{Co-doped} & \multicolumn{1}{ |c| }{Ni-doped} \\ \hline
 & $a$(\%) & -2.17 & -0.14 & -3.27 & -3.51\\ 
P-Se-terminated & $b$(\%) & -0.89 & -1.26 & +1.59 & +1.53 \\
 & $t$(\%) & +0.79 & +0.83 & +0.99 & +0.86 \\\hline
 & $a$(\%) & +2.45 & +3.36 & +2.45 & +1.45\\ 
Se-terminated & $b$(\%) & +3.29 & +4.28 & +3.64 & +4.95 \\
 & $t$(\%) & +7.06 & +7.85 & +8.80 & +8.25 \\\hline
 & $a$(\%) & -0.71 & -1.24 & -0.72 & -4.00\\ 
Zipper & $b$(\%) & +0.21 & +0.52 & -0.31 & +1.43 \\
 & $t$(\%) & +0.48 & -0.93 & +1.23 & +1.59\\\hline
 & $a$(\%) & -0.66 & -0.05 & -1.03 & -1.58\\ 
One-Se-plane & $b$(\%) & -0.65 & -0.05 & -1.05 & -1.57 \\
 & $t$(\%) & +2.82 & +3.60 & +4.98 & +5.16\\\hline
\end{tabular}
\caption{Percentage of expansion (+)/compression (-) for the lattice parameters $a$, $b$ and the total thickness $t$, after TM doping with respect to their pristine MLs.}
\label{tab:TM-doped}
\end{center}
\end{table*}

The Bader analysis for every structure shows appreciable differences between each model. Starting with the reconstructed Se-terminated ML, the valence of the Bi atoms located at the bottom plane has changed, i.e. the bismuth atoms at the upper Bi plane behave like Bi$^{+3}$, the regular valence for bismuths in bulk; while those at the bottom Bi plane behave like Bi$^{+2}$, which seems a particular feature for this reconstructed ML. Moreover, after the passivation of the reconstructed Se-terminated ML, the system returns to the original Se-terminated passivated ML, exhibiting identical electrical and structural properties, that is, the valence of Bi remains at its regular value, now again because hydrogens have avoided the electronic transference.
The results of the Bader analysis for the Zipper ML indicate that the two valences of Bi alternate between the atoms located at the same plane, whereas, for the One-Se-plane ML, the values remain like in the reconstructed Se-terminated ML, having the bottom Bi plane a higher value of charge (lower valence) than the upper one.

\subsection{Dopant Impurities (transition metal atoms)}
\subsubsection{Structural Properties}
\begin{figure}[h]
\centering
\includegraphics[width=0.48\textwidth]{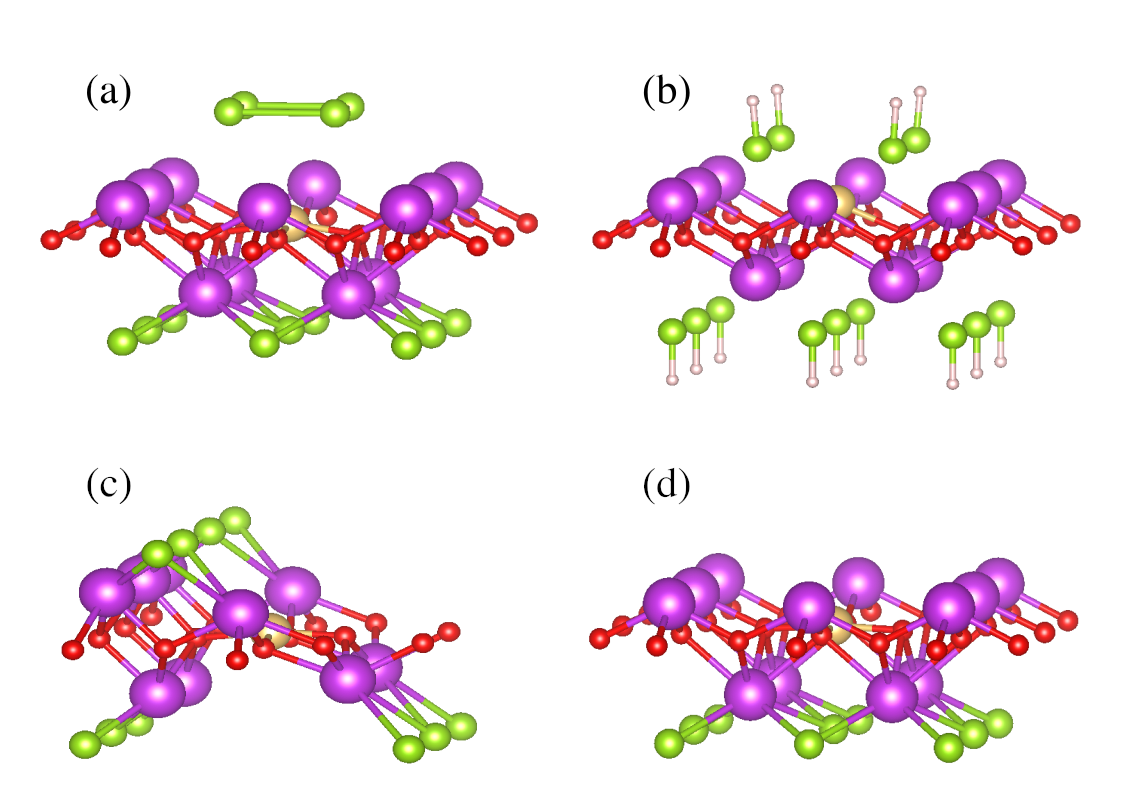}
 \caption{Structure of Mn-doped MLs. (a) Se-terminated ML (b) Passivated Se-terminated ML (c) Zipper ML (d) One-Se-plane ML}
 \label{fig:TM-dopedsymmetric}
\end{figure}

To introduce a magnetic moment in our systems, we have chosen Mn, Fe, Co and Ni as magnetic dopants. After a full relaxation of the Se-terminated ML pristine, remarkable structural changes are observed, as shown in Figure \ref{fig:TM-dopedsymmetric}a. The upper Se atoms have adopted a planar rectangular configuration with a side size of 2.26 Å on the $x$-axis and 2.97 Å on the $y$-axis, which has also been displaced upwards along the $z$-axis. This is a substantial change considering that the distances between the equivalent Se atoms at the bottom plane are 3.88 Å on the $x$-axis and 3.92 Å on the $y$-axis, respectively. The lattice parameters $a$ and $b$ of the Se-terminated ML (which, with doping, coincides with the R-Se-terminated ML) are always expanded independently of the nature of the dopant (Table \ref{tab:TM-doped}).

Moreover, while this monolayer presents no impurities, the upper and lower planes of the Bi and Se atoms result equidistant from the O atoms plane (which can be taken as a reference); and, after introducing the magnetic dopant that replace the Bi atom, a loss of symmetry along the z-axis is observed, causing a considerable increase in the thickness of the ML.
In addition, both, the TM atoms and the upper Bi plane are attracted towards the O plane, and the lower plane of the Bi atoms appears to be repelled respect to this plane. There are also subtle differences depending on the nature of the dopant. 
For example, as the dopant increases in electrons and decreases in size (Mn, Fe, Co and Ni, in this order), the ML thickness progressively changes by 0.74\%, 0.88\% and -0.50\% with respect to the previous dopant atom. This is due to the different effect of the electronic repulsion introduced by each atomic species.

Upon passivation of the Se-terminated ML (Fig. \ref{fig:TM-dopedsymmetric}b) the doped ML maintains better the structural characteristics of the pristine Se-terminated ML, avoiding the formation of the cluster of Se atoms at the top, with the exception of the dopant atoms, which accommodate falling towards the plane of oxygens (a general trend in all the studied MLs) as decreasing its size. In this sense, we could say that passivation provides structural stability to the system by maintaining symmetry along the z-axis. 
The thickness variations of the ML show similar behavior as the Se-terminated, to a lesser degree, and the changes of the lattice parameters result in either a compression (all $a$ and $b$ parameters of Mn- and Fe-doped MLs) or an expansion ($b$ parameter of Co- and Ni-doped MLs) (Table \ref{tab:TM-doped}).

Doping the Zipper ML (Fig. \ref{fig:TM-dopedsymmetric}c) leads only to small distortions of the system, such as a compression of the lattice parameter $a$, and an expansion of $b$ (see table \ref{tab:TM-doped}). However, symmetry along the perpendicular axis of the ML is still preserved. The total thickness of this ML decreases slightly under Fe doping, while there are slight thickness increases when other TM atoms are used as dopants. 

Finally, for the One-plane-Se ML, an asymmetry is again observed regarding the central O plane (Fig. \ref{fig:TM-dopedsymmetric}d). As mentioned above, this is due to the change in the valence of the lower bismuth atoms, which produces a different Coulomb repulsion between the upper and lower planes. The $a$ and $b$ parameters are compressed concerning their corresponding structure without dopant.
An increase in thickness is observed with respect to the pristine ML for each impurity used as dopant. 

\begin{table*}[ht]
\begin{center}
\resizebox{17.5cm}{!} {
\begin{tabular}{|c|c|c|c|c|c|c|}
\hline
\multicolumn{1}{ |c| }{Model of ML} & 
\multicolumn{1}{ |c| }{Magnitude} &
\multicolumn{1}{ |c| }{Mn-doped} &
\multicolumn{1}{ |c| }{Fe-doped} &
\multicolumn{1}{ |c| }{Co-doped} &
\multicolumn{1}{ |c| }{Ni-doped}
\\ \hline
Se-terminated & Cohesive energy (eV) & -2.417 & \bf-2.429  & -2.404 & -2.373 \\
 & MM(TM-atom)($\mu_B$) & 3.96 & 4.10 & 2.71 & 1.67 \\
 & Total MM ($\mu_B$) & 4.68 & 5.98 & 4.80 & 2.86 \\
 & Electron effective mass (m$_{e}$) & - & - & - & - \\ \hline
 P-Se-terminated & Cohesive energy (eV) & -1.88 & \bf-1.89 & -1.87 & -1.84 \\
 & MM(TM-atom)($\mu_B$) & 4.72 & 3.71 & 2.74 & 1.69 \\
 & Total MM ($\mu_B$) & 4.51 & 3.83 & 2.96 & 2.13 \\
 & Electron effective mass (m$_{e}$) & - & - & - & - \\ \hline
Zipper & Cohesive energy (eV) & \bf-2.432 & -2.428 & -2.39 & -2.36 \\
 & MM(TM-atom)($\mu_B$) & 3.99 & 4.19 & 3.04 & 1.70 \\
 & Total MM ($\mu_B$) & 4.00 & 5.00 & 4.00 & 2.08 \\ 
 & Electron effective mass (m$_{e}$) & -0.892 (1.397) & -0.933 (21.815) & -0.938 (1.571) & - \\ \hline
One-Se-plane & Cohesive energy (eV) & -2.377 & \bf-2.405 & -2.374 & -2.291 \\
 & MM(TM-atom)($\mu_B$) & 3.90 & 4.07 & 2.69 & 1.29 \\
 & Total MM ($\mu_B$) & 4.00 & 5.00 & 3.23 & 2.54 \\ 
 & Electron effective mass (m$_{e}$) & -0.204 (2.376) & -0.234 (0.433) & - & - \\ \hline
\end{tabular}}
\caption{Cohesive energies (eV), magnetic moment (MM) ($\mu_B$) per TM-atom, total magnetic moment (total MM) ($\mu_B$) of the ML and electron effective mass (m$_{e}$) in B.V. (B.C.). }
\label{tab:data-doped}
\end{center}
\end{table*}

The cohesive energies of doped monolayers are reported in Table \ref{tab:data-doped}. It is observed that the most stable non-stoichiometric structures are those with Fe as an impurity, whereas The Zipper monolayer is most stable doped with Mn.
The Mn-doped Zipper monolayer has the highest cohesive energy (-2.432 eV), even higher than the Se-terminated ML (-2.429 eV) and the One-Se-plane ML (-2.405 eV) doped with Fe.

\subsubsection{Magnetic Properties}

The magnetism induced by doping with 3d TM atoms comes mainly from the d electrons of the magnetic impurity used to replace the Bi atoms. In the case of the Se-terminated ML, there is also a small contribution coming from the induced polarization of some Se atoms. This magnetic polarization appears due to an asymmetric charge transfer in the different spin channels.  The magnetic configuration adopted by the upper plane of Se atoms concerning the direction of the magnetic moment of the TM atom is ferromagnetic (FM), as can be seen in the Table \ref{tab:not_passivated-ML}. The largest magnetic moment is achieved with Fe impurities (6.0 $\mu_B$) and it is originated from a FM configuration among the Fe atom (with 4.1 $\mu_B$) and both the Se atoms of the upper plane and those Se atoms closest to the impurity (located in the lower plane). There is also magnetic polarization of the O atoms surrounding the TM atom, as well as the Se atoms in the lower layer that are second neighbors to the impurity, with $\sim$0.1 $\mu_B$ each. The value of 4.7 $\mu_B$ for Mn doping is explained because the Mn atom couples with the Se atoms at the upper plane through a FM state; while the Se atoms at the bottom plane are antiferromagnetic AF (see Table \ref{tab:not_passivated-ML}). On the other hand, when this type of ML is doped with Co, FM polarization of both upper and bottom Se atoms adopt the same order of magnitude ($\sim$0.2 $\mu_B$), which leads to an important increase of the total magnetic moment in comparison with the Co atom (4.8 $\mu_B$ y 2.7 $\mu_B$, respectively, Table \ref{tab:data-doped}).This effect is not so relevant when Ni act as dopant because the polarization only happens to the upper Se atoms. 

In the P-Se-terminated case, Mn atoms have an AF coupling with the Se atoms in the upper plane (which are close to them), which reduces the total magnetic moment (4.5 $\mu_B$) with respect to the value of the TM atom (4.7 $\mu_B$, see Table \ref{tab:data-doped}). On the other hand, Fe, Co and Ni atoms slightly polarize, towards their neighboring atoms, with a FM coupling, which enhances their total magnetic moment. 

For the Zipper ML case (Table \ref{tab:Zipper-Magnetics-moments}), the Fe, Co and Ni atoms help to polarize magnetically the oxygen atoms, instead of the Se atoms, with a FM coupling among them. This increases the total magnetic moment of the system. In the particular case of Mn-doping (which has the highest cohesive energy in this type of ML), no appreciable magnetic polarization is observed in its neighborhood. Finally, in the One-Se-plane ML (Table \ref{tab:One-Se-plane}), magnetic polarization is induced in all the Se atoms when doped with Co and Ni; while it takes place in a single Se atom (this located immediately under the impurity) when doped with Mn and Fe. In this last case (Fe-doped) a polarization of oxygen atoms is also observed.

Thus, structure strongly influences the total magnetic moment, i.e., the interplay between the structural distortions and the asymmetric charge transfer occurring at each structure plays a key role in determining the total magnetic moment of the ML.

\subsubsection{DOS and Band Structures}

Figure \ref{DOSTM} shows the DOS for the TM-doped monolayers. It reveals that both Se-terminated and passivated MLs are metallic regardless of the TM dopant considered. 

For the Se-terminated MLs, the Fe-doped (fig. \ref{DOSTM}(a2)) has the biggest number of states at the Fermi level (sum of the contributions of the spins above and below), and according to the Drude's model, it would have a higher electrical conductivity. The electrical conductivity decreases when the Se-terminated ML is doped with Ni, Co and Mn (Figs. S\ref{fig_dos-mono}(a4), S\ref{fig_dos-mono}(a3) and S\ref{fig_dos-mono}(a1), respectively). For the doped P-Se-terminated MLs (figs. S\ref{fig_dos-mono}(b)), the semiconductor character exhibited in the pristine case is lost, and  a progressively decreasing electrical conductivity is obtained when doping Ni, Co, Fe and Mn, in this order.

On the other hand, the electrical character of the stoichiometric MLs changes depending on the doping TM atom. The Zipper ML (figs. S\ref{fig_dos-mono}(c)) has a semiconductor behavior when doped with Mn, Fe, or Co with bandgaps of 1.75, 0.96 and 0.76 eV, respectively, and it behaves as a metal when doping with Ni. The One-Se-plane monolayers (figs. S\ref{fig_dos-mono}(d)) doped with Mn and Fe are semiconductors with 0.36 and 0.56 eV bandgaps, while the ones doped with Co and Ni are metallic with a higher conductivity in the Ni-doped case.

\begin{figure*}[]
\includegraphics[width=0.3\textwidth]{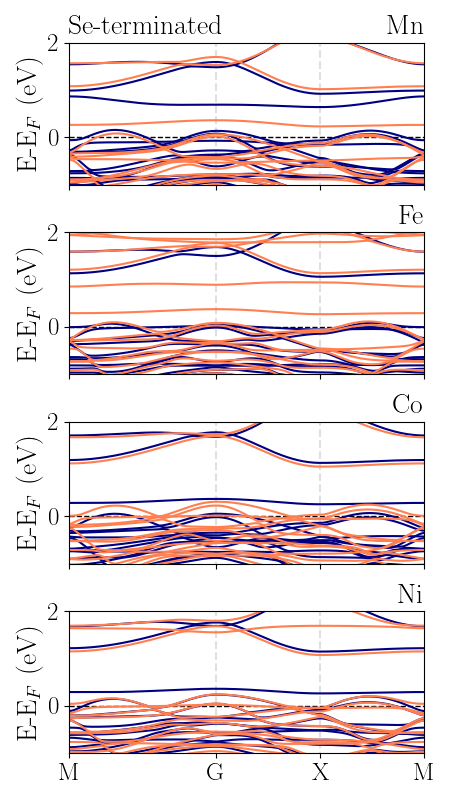}\includegraphics[width=0.3\textwidth]{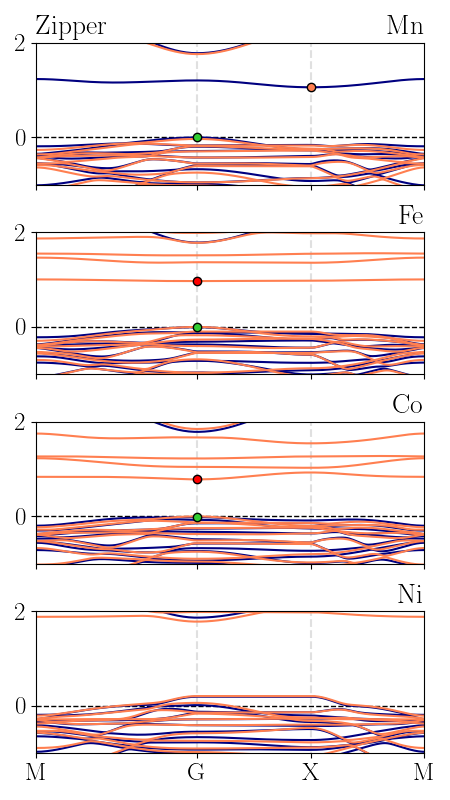}\includegraphics[width=0.3\textwidth]{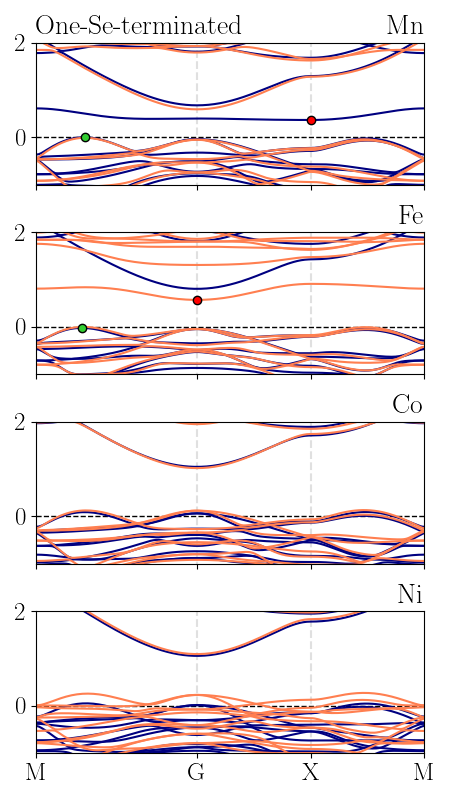}
 \caption{Band structure for the TM-doped reconstructed Se-terminated, Zipper and One-Se-terminated MLs: (a) Mn-dopant (b) Fe-dopant (c) Co-dopant (d) Ni-dopant. The majority (blue lines) and minority (red lines) bands are shown.}
 \label{fig:BandDopedSe}
\end{figure*}

The band structures of the doped systems (see Figure \ref{fig:BandDopedSe}) show that the introduction of impurities into the systems results in the appearance of localized states of low dispersion in the band structure. These states are located in different energetic positions depending on the type of monolayer and the dopant. Such bands come from the d states of the impurity that sometimes hybridize with atoms of the monolayer. 
In particular, in the Se-terminated monolayer with Mn, a hybridization is observed between the d orbitals of Mn and the p states of O. This gives rise to an impurity band located in the gap between the last half-occupied and the empty bands of the pristine R-Se-terminated monolayer; which is located at 0.69 eV and composed of spins up. Another localized band is observed at 0.3 eV above E$_F$, and it is composed by the Se atoms that form the tetramer. Due to the FM coupling among these atoms and the Mn atom, this localized band is composed by unoccupied p states that would be occupied by spin-down electrons in excited states. However, when the dopant is Fe, the impurity band does not hybridize with the atoms in the pristine monolayer, and an energy shift towards higher values (around 0.9 eV) takes place. 
Conversely, in the Co and Ni cases, the impurity band is mixed with the unoccupied band of the monolayer, whereas the localized band, which is formed by the Se tetramer, is composed by unoccupied spin-up states, due to the existing AF coupling with the dopant atoms.

In the passivated ML, no impurity bands are observed within the bandgap, again, due to the presence of hydrogen atoms that eliminates the unsaturated bonds. In the band structure of the stoichiometric MLs, doping with Mn atoms leads to impurity bands in the bandgap (at 1.16 eV for Zipper and 0.88 eV for One-Se-plane with respect to the maximum of the valence band), which would be occupied with up-spin electrons, as a result of the hybridization between Mn and O atoms (in the One-Se-plane ML there is also an important contribution from the Bi atoms). With the Fe atom as dopant, there are three impurity bands in the Zipper ML, with spins down, located at 0.98, 1.38 and 1.51 eV in the bandgap, and only one impurity band in the One-Se-plane ML, which is located at 0.84 eV in the bandgap and correspond to spin down electrons as well. With a Co atom, the Zipper ML presents impurity bands with spins down at 0.82, 1.04, 1.24 and 1.66 eV, and no impurity bands are observed in the bandgap for the One-Se-plane ML. Similarly, by using Ni as a dopant, there is only one spin-down impurity band in the bandgap, at 1.87 eV, but none is observed in the bandgap of the One-Se-plane ML band structure.

The Bader analysis allows to confirm again that for the Se-terminated ML, after doping with the TM atoms, the valence of the bottom Bi ions remains as Bi$^{+2}$ and the upper Bi ions behave as Bi$^{+3}$ (as we mentioned above, the regular value before reconstruction occurs). In analogy to the Se-terminated ML without dopant, the Se upper ions show null valence, whereas bottom Se ions act with valence -1. Oxygen atoms behave as O$^{-2}$, similarly to the pristine structure, and, in general, the TM dopants do loose charge, which is transferred towards the pristine ML. 

The charge distribution in the Zipper ML is analogous to its corresponding system with no dopant, that is, the two valence values exhibited by Bi alternate between ions in the same plane (Fig. \ref{fig:valenceZ}) being the valences of Se and O atoms, -1 and -2, respectively. Moreover, in the One-Se-plane ML, O and Se behave in similar way to the not doped case (valence -2 and -1); and only Bi ions present changes depending the plane they are located at (i.e., bottom plane as Bi$^{+2}$ and upper plane as Bi$^{+3}$). Such changes in the behavior of the valence would also explain the structural transformations induced by the introduction of the impurities.

\section{Conclusions}

We present a detailed study of the structural, electronic and magnetic properties of four different 
Bi$_2$O$_2$Se-based MLs. Two of them are stoichiometric and two are not. 
One aim is to evaluate their suitability for modeling purposes. 
We have observed that, except for the passivated ML, those systems with excess Se present a metallic character; which confirms the sensitivity of these systems to small changes in the carrier concentration.
Although the passivated model has predominated traditionally among theoretical studies, it is worth 
noticing the lack of coherence since, for experiments, most of the samples are grown via CVD, 
which does not involve hydrogen. 
On the other hand, the stoichiometric monolayers retained the semiconductor bulk behavior, although only the Zipper ML reproduces successfully the observed thickness dependence of the gap. This ML presents the closest gap values to the experimental measurements, as well as a very high cohesion energy, which can be of interest for further technical developments in fields such as 
optics, thermal properties and thermoelectricity. 
No imaginary frequencies were obtained for the vibrational modes at the $\Gamma$ point (for none of the monolayers studied) which indicates their structural stability, and experimental feasibility.  
The four models considered were doped with TM atoms (Mn, Fe, Co and Ni) in order to study each monolayer responses under the presence of magnetic species, with the aim to explore new exciting applications coming from the combined effect of magnetism in 2D systems and the thermoelectric and optic properties of the BiSeO-based samples. Indeed, the "extra" electrons, coming from the TM dopants, caused the appearance of delocalized states in the gap (coming from their d orbitals), which led either to narrow the gap of the semiconducting monolayers or to their transition towards a metallic behavior. Finally, from the magnetic viewpoint, it is worth noticing that, in some of the presented models, the impurities help to polarize both the surrounding oxygens and the selenium atoms. This can result in a remarkable increase of the total magnetic moment of the monolayer (for instance 
the Fe-doped Se-terminated ML, which exhibits up to 6 $\mu_B$/U.C.), what makes these 2D materials very interesting for spintronic purposes as well.\\

\section{Conflicts of interest}
There are not conflicts to declare

\section{Acknowledgements}

This work has been supported by the project ANID/CONICYT FONDECYT Grant No. 1210193 ``Financiamiento basal para centros científicos y tecnológicos de excelencia AFB180001"
We are also very grateful to Prof. Johan Mazo-Zuluaga from the University of Antioquia for his help, advice and orientation with the manuscript.

\bibliographystyle{unsrtnat}
\bibliography{References}

\pagebreak

\newpage

\onecolumn
\begin{center}

\textbf{\large Supplemental Material. Structural design of Bi$_2$O$_2$Se monolayers:  electronic and magnetic properties with transition metal (Mn, Fe, Co, Ni) doping}

\vspace{0.5cm}
I. Arias$^{1,2}$, A.M León$^1$, J. Mejía-López$^1$

$^1$Facultad de Física, Pontificia Universidad Católica de Chile, Santiago, Chile.\\
$^2$ Institute of Experimental Physics, University of Warsaw

\end{center}

\setcounter{figure}{0} 
\setcounter{section}{0} 
\setcounter{equation}{0}
\setcounter{table}{0}
\setcounter{page}{1}
\renewcommand{\thepage}{S\arabic{page}} 
\renewcommand{\thesection}{S\Roman{section}}   
\renewcommand{\thetable}{S\arabic{table}}  
\renewcommand{\thefigure}{S\arabic{figure}} 
\renewcommand{\theequation}{S\arabic{equation}} 

\begin{figure*}[h]
\centering
\includegraphics[width=0.50\textwidth]{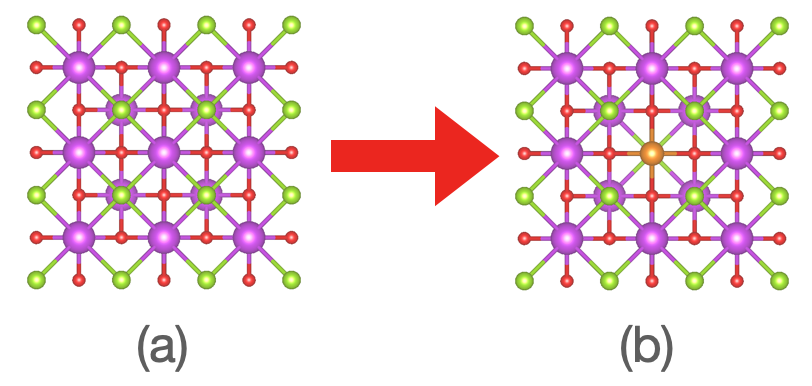}
 \caption{Top view of ML: (a) before and (b) after doping.}
 \label{fig:Doping}
\end{figure*}

Figure \ref{fig:Doping} shows a BOS supercell doped with a transition metal (12.5\%) for a Se-terminated structure. The doping of the other monolayers studied in this work is carried out similarly.

\begin{figure*}[h]
\centering
{\includegraphics[width=0.5\textwidth]{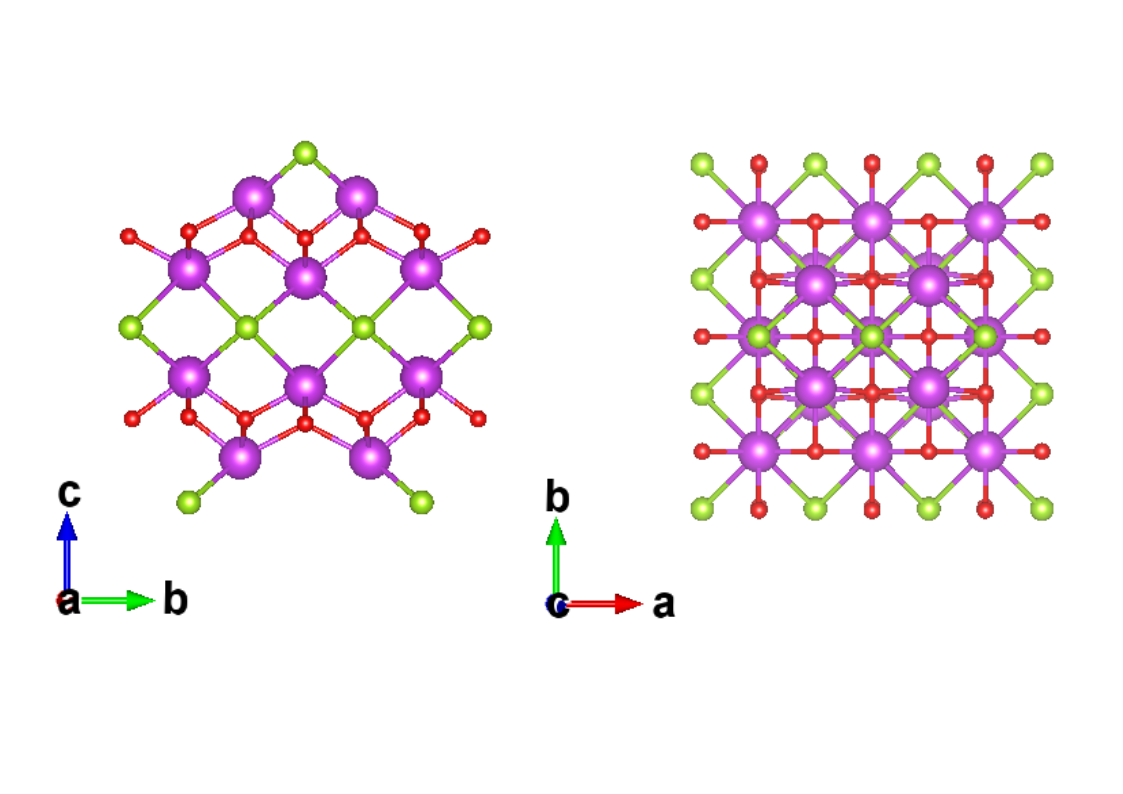}}
\caption{Structure of a bilayer of Bi$_2$O$_2$Se: side (a) and top (b) views.}
\label{fig:ZipperBL}
\end{figure*}

Figure \ref{fig:ZipperBL} shows the bilayer model for Bi$_2$O$_2$Se. This structure is built from two Zipper monolayers to keep the bulk symmetry \cite{Wei2019}.

\begin{figure*}[h]
\centering
\includegraphics[width=0.6\textwidth]{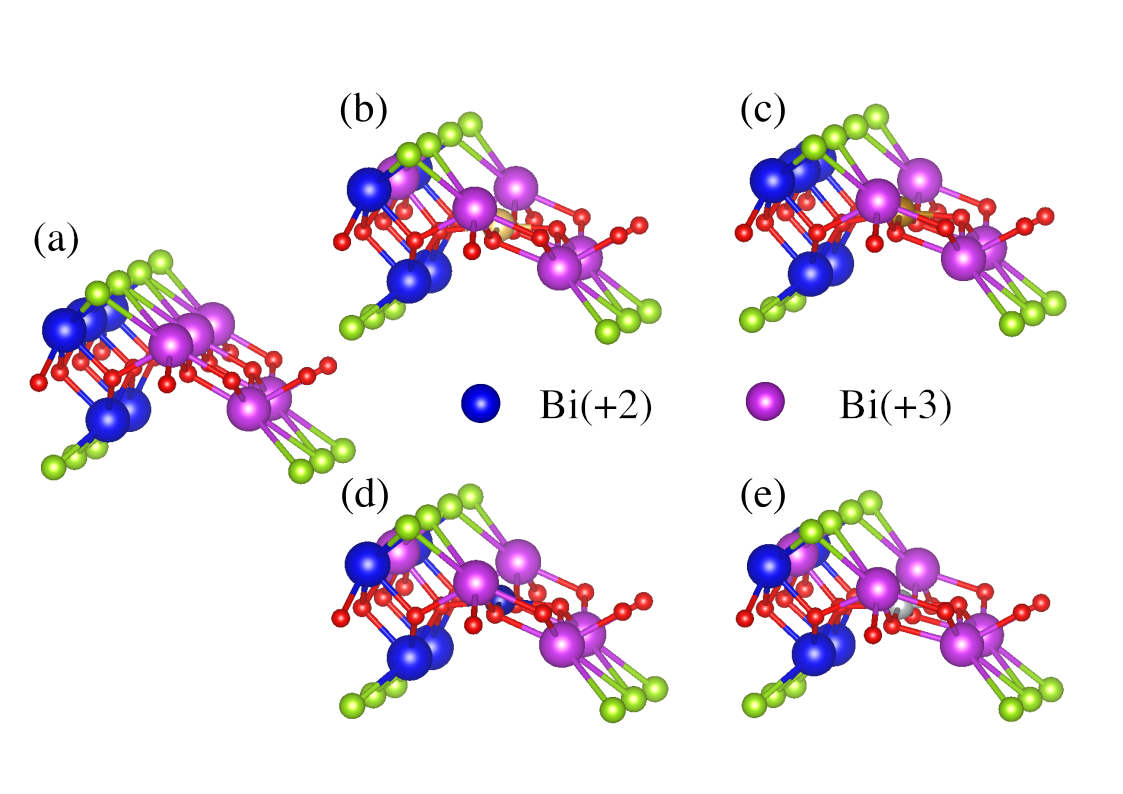}
 \caption{Arrangement of bismuth valence values for the Zipper ML: (a) Pristine Zipper ML (b) Mn-doped (c) Fe-doped (d) Co-doped (e) Ni-doped. Blue spheres represent Bi$^{+2}$ ions and violet spheres represent Bi$^{+3}$ ions.\\}
 \label{fig:valenceZ}
\end{figure*}

\begin{figure*}[ht]
\includegraphics[width=0.245\textwidth]{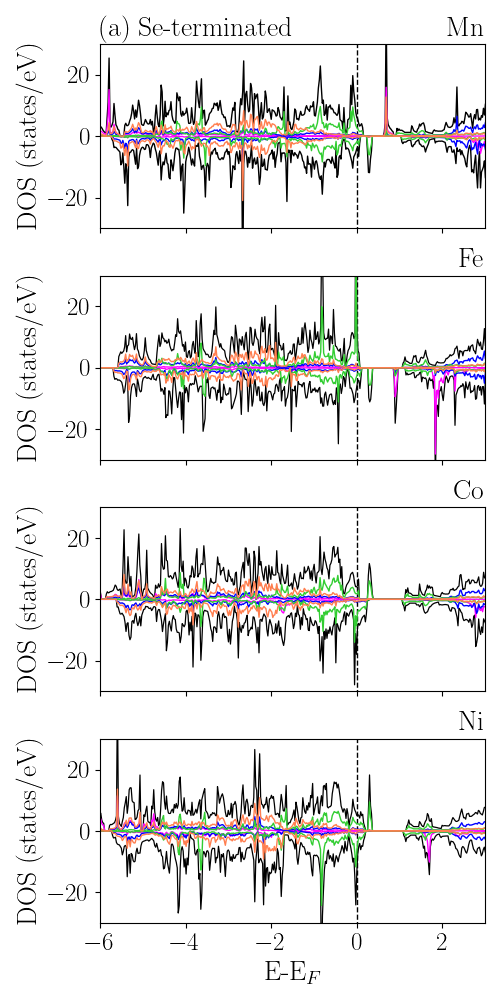}\includegraphics[width=0.245\textwidth]{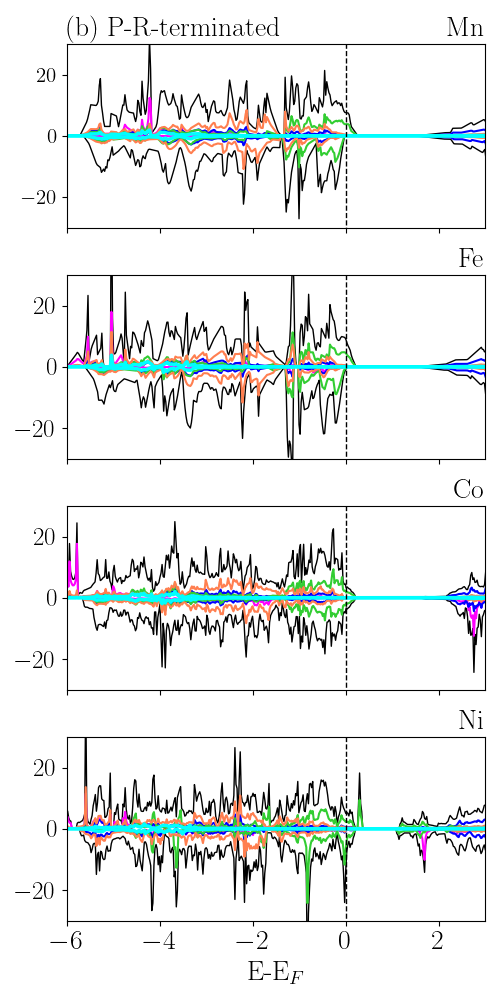}
\includegraphics[width=0.245\textwidth]{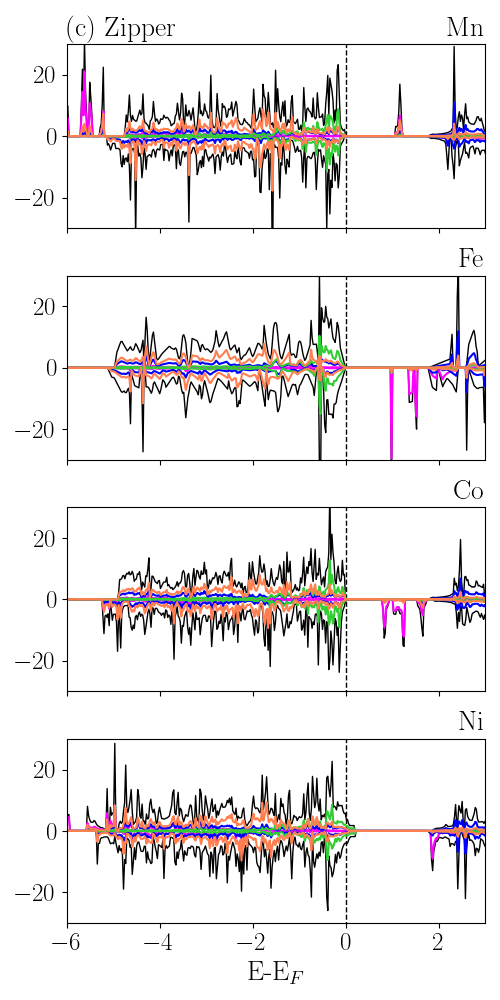}\includegraphics[width=0.245\textwidth]{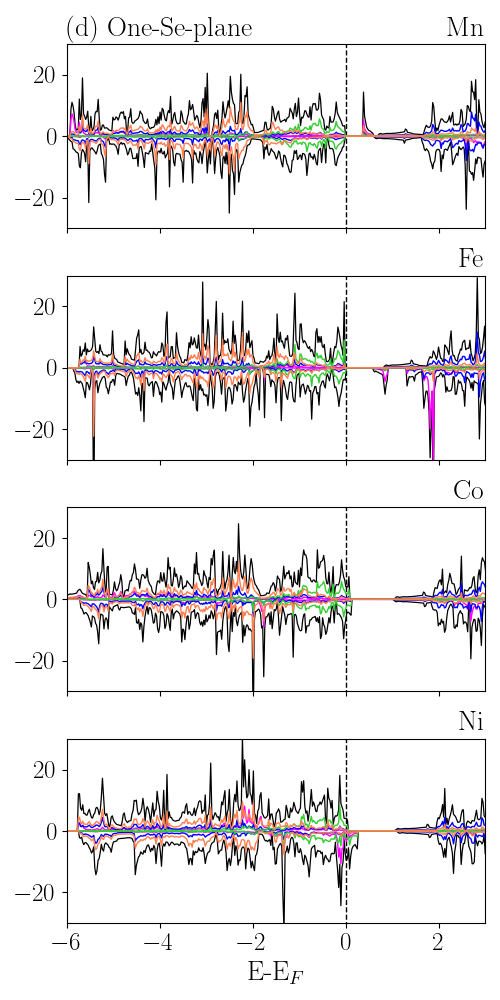}
\includegraphics[width=0.7\textwidth]{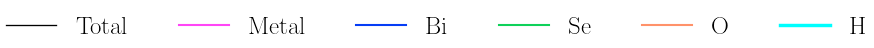}
\caption{Total and partial density of states of TM-doped. (a) Se-terminated ML, (b) P-Se-terminated ML, (c) Zipper ML and (d) One-Se-plane.}
\label{DOSTM}
\end{figure*}

Tables \ref{tab:not_passivated-ML},  \ref{tab:passivated}, \ref{tab:Zipper-Magnetics-moments}, and \ref{tab:One-Se-plane} shows the magnetic moment for Se-terminated, P-Se terminated, Zipper and One-Se plane structures, respectively. Here, we can observe 
that the structure strongly influences the total magnetic moment; this is due to the strong interplay between the structural distortions present in the different ML models and the asymmetric charge transfer occurring at each structure plays a key role in determining the total magnetic moment of the ML.\\

\begin{table*}[ht]
\begin{center}
\begin{tabular}{| c | c | c | c | c | c | c | c | c | c | c |}
\hline
\multicolumn{11}{ |c| }{Magnetic moments in the Se-terminated ML ($\mu_B$)} \\ \hline
 \multicolumn{1}{|c|}{TM-atom} & \multicolumn{4}{c|}{Bottom plane} & \multicolumn{4}{c|}{Upper plane} & \multicolumn{1}{c|}{Dopant} & \multicolumn{1}{c|}{Total} \\
 \multicolumn{1}{|c|}{} & \multicolumn{1}{c}{Se1} & \multicolumn{1}{|c}{Se2} & \multicolumn{1}{|c}{Se3} & \multicolumn{1}{|c|}{Se4} & \multicolumn{1}{c}{Se5} & \multicolumn{1}{|c}{Se6} & \multicolumn{1}{|c}{Se7} & \multicolumn{1}{|c|}{Se8} & \multicolumn{1}{c|}{} & \multicolumn{1}{c|}{} \\ \hline
 Mn & -0.1 & -0.1 & -0.1 & 0.0 & 0.2 & 0.2 & 0.2 & 0.2 & 4.0 & 4.7 \\ 
Fe & 0.0 & 0.1 & 0.1 & 0.3 & 0.2 & 0.2 & 0.2 & 0.2 & 4.1 & 6.0\\
Co & 0.2 & 0.3 & 0.2 & 0.2 & 0.2 & 0.2 & 0.2 & 0.2 & 2.7 & 4.8\\
Ni & 0.0 & 0.0 & 0.0 & 0.1 & 0.2 & 0.2 & 0.2 & 0.2 & 1.7 & 2.9\\ \hline
\end{tabular}
\caption{Magnetic moments in the Se-terminated monolayer ($\mu_B$).}
\label{tab:not_passivated-ML}
\end{center}
\end{table*}



\begin{table*}[ht]
\begin{center}
\begin{tabular}{| c | c | c | c | c | c | c | c | c | c | c |}
\hline
\multicolumn{11}{ |c| }{Magnetic moments in the P-Se-terminated ML ($\mu_B$)} \\ \hline
 \multicolumn{1}{|c|}{TM-atom} & \multicolumn{4}{c|}{Bottom plane} & \multicolumn{4}{c|}{Upper plane} & \multicolumn{1}{c|}{Dopant} & \multicolumn{1}{c|}{Total} \\
 \multicolumn{1}{|c|}{} & \multicolumn{1}{c}{Se1} & \multicolumn{1}{|c}{Se2} & \multicolumn{1}{|c}{Se3} & \multicolumn{1}{|c|}{Se4} & \multicolumn{1}{c}{Se5} & \multicolumn{1}{|c}{Se6} & \multicolumn{1}{|c}{Se7} & \multicolumn{1}{|c|}{Se8} & \multicolumn{1}{c|}{} & \multicolumn{1}{c|}{} \\ \hline
 Mn & - & - & - & - & -0.1 & -0.1 & -0.1 & -0.1 & 4.7 & 4.5 \\ 
Fe & - & - & - & - & - & - & - & - & 3.7 & 3.8\\
Co & - & - & - & - & - & - & - & - & 2.7 & 3.0 \\
Ni & - & - & - & - & - & - & - & - & 1.7 & 2.0\\ \hline
\end{tabular}
\caption{Magnetic moments in the P-Se-terminated monolayer ($\mu_B$).}
\label{tab:passivated}
\end{center}
\end{table*}

\begin{table*}[ht]
\begin{center}
\resizebox{13.5cm}{!} {
\begin{tabular}{|c|c|c|c|c|c|c|}
\hline
\multicolumn{7}{ |c| }{Magnetic moments in the Zipper TM-doped monolayer} \\ \hline
Dopant & O3 & O4 & O6 & O8 & TM-atom ($\mu_B$) & Total magnetic moment ($\mu_B$) \\ \hline
Mn & - & - & - & - & 4.0 & 4.0\\
Fe & 0.1 & 0.1 & 0.2 & 0.2 & 4.2 & 5.0 \\
Co & 0.2 & 0.2 & 0.2 & 0.2 & 3.0 & 4.0 \\
Ni & - & 0.1 & 0.1 & 0.1 & 1.7 & 2.1 \\ \hline
\end{tabular}}
\caption{Magnetic moments in the Zipper TM-doped monolayer ($\mu_B$).}
\label{tab:Zipper-Magnetics-moments}
\end{center}
\end{table*}

\begin{table*}[ht]
\begin{center}
\resizebox{18cm}{!} {
\begin{tabular}{|c|c|c|c|c|c|c|c|c|c|c|}
\hline
\multicolumn{11}{ |c| }{Magnetic moments in One-Se-plane TM-doped ML} \\ \hline
Dopant & Se1 & Se2 & Se3 & Se4 & O3 & O4 & O6 & O8 & TM-atom ($\mu_B$) & Total monolayer ($\mu_B$) \\ \hline
Mn & - & - & - & 0.1 & - & - & - & - & 3.9 & 4.0\\
Fe & - & - & - & 0.3 & 0.1 & 0.1 & 0.1 & 0.1 & 4.1 & 5.0 \\
Co & 0.1 & 0.1 & 0.1 & 0.1 & - & - & - & - & 2.7 & 3.2 \\ 
Ni & 0.3,0.1 & 0.2,0.1 & 0.2,0.1 & 0.2 & 0.1,0 & 0.1,0 & 0.1,0 & 0.1,0 & 1.3,1.7 & 2.5,2.3  \\ \hline
\end{tabular}}
\caption{Magnetic moments in the One-Se-plane TM-doped ML.}
\label{tab:One-Se-plane}
\end{center}
\end{table*}

Figure \ref{DOSTM} displays the Total and partial density of states for the TM-doped structures in the valence band. As a general characteristic,  Se atoms contribute mainly around the Fermi level, and O atoms are between -5 to -2 eV, while the metal atoms contribute mostly in the conduction band. Furthermore, we have found that independent of the metal, the Se-terminated ML (Fig.\ref{DOSTM}(a)) remains in a metic state; however, to passivate this structure (Fig.\ref{DOSTM}(b)), the system can exhibit a semimetallic or metallic character under Mn-Fe and Ni-Co dopping, respectively. In the case of Zipper and One-Se-plane (Fig.\ref{DOSTM}(c)-(d)), the system can keep its semiconductor state or go toward a metallic state depending on the metal: Zipper only displays one metallic state under Ni doping and One-Se-plane under Ni and Co doping.\\

Table \ref{tab:Bader1} displays the Bader analysis for Bi, Se, O, and H atoms for each proposed monolayer model. Here we can observe that the valence of Se decreases for the passivated system due to the electron transference to H atoms. Furthermore, the Zipper structure gets a Bi mixed valence character with Bi$^{+2}$ and Bi$^{+3}$. Fig. \ref{fig:valenceZ} shows the valence rearrangement (Bi$^{+2}$ and Bi$^{+3}$) for this structure.\\

\begin{table*}[h]
\begin{center}
\begin{tabular}{| c | c | c | c | c | c | c |}
\hline
\multicolumn{7}{ |c| }{Charge of ions (valence)} \\ \hline
 \multicolumn{1}{|c|}{Type of monolayer} & \multicolumn{2}{c|}{Bi} & \multicolumn{2}{c|}{Se} & \multicolumn{1}{c|}{O} & \multicolumn{1}{c|}{H}\\
 \multicolumn{1}{|c|}{} & \multicolumn{1}{c|}{Top} & \multicolumn{1}{c|}{Bottom} & \multicolumn{1}{c|}{Top} & \multicolumn{1}{c|}{Bottom} & \multicolumn{1}{c|}{} & \multicolumn{1}{c|}{}\\ \hline
Bulk & 2.2 (+3) & 2.2 (+3) & 7.0 (-1) & 7.0 (-1) & 8.3 (-2) & - \\
Se-terminated & 2.2 (+3) & 2.2 (+3) & 6.5 (-1) & 6.5 (-1) & 8.2 (-2) & - \\
P-Se-terminated & 2.2 (+3) & 2.2 (+3) & 5.5 (0) & 5.5 (0) & 8.3 (-2) & 2.0 (-1)\\
R-Se-terminated & 2.1 (+3) & 3.3 (+2) & 6.2 (0) & 6.7 (-1) & 7.8 (-2) & -\\
Zipper & 3.1(+2)/2.1(+3) & 3.1(+2)/2.1(+3) & 6.9 (-1) & 6.9 (-1) & 7.9 (-2) & -\\ 
One-Se-plane & 2.1(+3) & 3.4(+2) & 6.8 (-1) & 6.8 (-1) & 7.8 (-2) & -\\
\hline
\end{tabular}
\caption{Bader analysis per atom: the values in parenthesis indicate the nominal valence of each ion. Coplanar bismuth ions alternate two valence values in the case of the Zipper ML.} 
\label{tab:Bader1}
\end{center}
\end{table*}

Tables \ref{tab:Se-terminated}, 
\ref{tab:P-Se-terminated}
\ref{tab:doped-Zipper} and \ref{tab:doped-One-Se-plane}, show the Bader analysis of charge for the Se-terminated, P-Se terminated, Zipper and One-Se plane ML structures, respectively. As a general behavior, we have evidenced that the charge valence of TM doped monolayers behave the same as the non-doped systems for all the structures. Furthermore, all the TM dopants transfer charge to the atoms of the pristine structure.

\begin{table*}[ht]
\begin{center}
\begin{tabular}{| c | c | c | c | c | c | c |}
\hline
\multicolumn{7}{ |c| }{Charge of ions (valence) at the Se-terminated ML} \\ \hline
 \multicolumn{1}{|c|}{Type of doped ML} & \multicolumn{2}{c|}{Bi} & \multicolumn{2}{c|}{Se} & \multicolumn{1}{c|}{O} & \multicolumn{1}{c|}{TM}\\
 \multicolumn{1}{|c|}{} & \multicolumn{1}{c|}{Top} & \multicolumn{1}{c|}{Bottom} & \multicolumn{1}{c|}{Top} & \multicolumn{1}{c|}{Bottom} & \multicolumn{1}{c|}{} & \multicolumn{1}{c|}{}\\ \hline
Mn-doped & 2.1 (+3) & 3.3 (+2) & 6.2 (0) & 6.8 (-1) & 7.7 (-2) & 5.2 \\
Fe-doped & 2.1 (+3) & 3.3 (+2) & 6.1 (0) & 6.8 (-1) & 7.7 (-2) & 6.3 \\
Co-doped & 2.1 (+3) & 3.2 (+2) & 6.1 (0) & 6.7 (-1) & 7.7 (-2) & 7.7 \\ 
Ni-doped & 2.1(+3) & 3.0(+2) & 6.1 (0) & 6.7 (-1) & 7.8 (-2) & 8.8 \\
\hline
\end{tabular}
\caption{Resulting Bader analysis after doping the Se-terminated monolayer.} 
\label{tab:Se-terminated}
\end{center}
\end{table*}

\begin{table*}[h]
\begin{center}
\begin{tabular}{| c | c | c | c | c | c | c |}
\hline
\multicolumn{7}{ |c| }{Charge of ions (valence) at the P-Se-terminated ML} \\ \hline
 \multicolumn{1}{|c|}{Type of doped ML} & \multicolumn{2}{c|}{Bi} & \multicolumn{2}{c|}{Se} & \multicolumn{1}{c|}{O} & \multicolumn{1}{c|}{TM}\\
 \multicolumn{1}{|c|}{} & \multicolumn{1}{c|}{Top} & \multicolumn{1}{c|}{Bottom} & \multicolumn{1}{c|}{Top} & \multicolumn{1}{c|}{Bottom} & \multicolumn{1}{c|}{} & \multicolumn{1}{c|}{}\\ \hline
Mn-doped & 2.2 (+3) & 2.2 (+3) & 5.4 (+1) & 5.5 (+1) & 8.2 (-2) & 5.5 \\
Fe-doped & 2.2 (+3) & 2.2 (+3) & 5.4 (+1) & 5.5 (+1) & 8.2 (-2) & 6.6 \\
Co-doped & 2.5 (+3) & 2.2 (+3) & 5.4 (+1) & 5.5 (+1) & 8.1 (-2) & 7.7 \\ 
Ni-doped & 2.4 (+3) & 2.1 (+3) & 5.4 (+1) & 5.5 (+1) & 8.1 (-2) & 8.7 \\
\hline
\end{tabular}
\caption{Resulting Bader analysis after doping the P-Se-terminated monolayer.} 
\label{tab:P-Se-terminated}
\end{center}
\end{table*}

\begin{table*}[ht]
\begin{center}
\begin{tabular}{| c | c | c | c | c | c | c |}
\hline
\multicolumn{7}{ |c| }{Charge of ions (valence) at the Zipper ML} \\ \hline
 \multicolumn{1}{|c|}{Type of doped ML} & \multicolumn{2}{c|}{Bi} & \multicolumn{2}{c|}{Se} & \multicolumn{1}{c|}{O} & \multicolumn{1}{c|}{TM}\\
 \multicolumn{1}{|c|}{} & \multicolumn{1}{c|}{Top} & \multicolumn{1}{c|}{Bottom} & \multicolumn{1}{c|}{Top} & \multicolumn{1}{c|}{Bottom} & \multicolumn{1}{c|}{} & \multicolumn{1}{c|}{}\\ \hline
Mn-doped & 2.1 (+3)/ 3.1 (+2) & 2.1 (+3)/ 3.1 (+2) & 7.4 (-1) & 6.9 (-1) & 7.8 (-2) & 5.2 \\
Fe-doped & 2.1 (+3)/ 3.1 (+2) & 2.1 (+3)/ 3.1 (+2) & 6.9 (-1) & 6.9 (-1) & 7.8 (-2) & 6.2 \\
Co-doped & 2.1 (+3)/ 3.1 (+2) & 2.1 (+3)/ 3.1 (+2) & 7.4 (-1) & 6.9 (-1) & 7.8 (-2) & 7.4 \\ 
Ni-doped & 2.1 (+3)/ 3.1 (+2) & 2.1 (+3)/ 2.6 (+2) & 7.3 (-1) & 6.8 (-1) & 7.9 (-2) & 8.7 \\
\hline
\end{tabular}
\caption{Resulting Bader analysis after doping the Zipper monolayer. Valences alternate into the same plane.} 
\label{tab:doped-Zipper}
\end{center}
\end{table*}

\begin{table*}[ht]
\begin{center}
\begin{tabular}{| c | c | c | c | c | c |}
\hline
\multicolumn{6}{ |c| }{Charge of ions (valence) at the One-Se-plane ML} \\ \hline
 \multicolumn{1}{|c|}{Type of doped ML} & \multicolumn{2}{c|}{Bi} & \multicolumn{1}{c|}{Se} & \multicolumn{1}{c|}{O} & \multicolumn{1}{c|}{TM}\\
 \multicolumn{1}{|c|}{} & \multicolumn{1}{c|}{Top} & \multicolumn{1}{c|}{Bottom} & \multicolumn{1}{c|}{} & \multicolumn{1}{c|}{} & \multicolumn{1}{c|}{}\\ \hline
Mn-doped & 2.1 (+3) & 3.4 (+2) & 6.9 (-1) & 7.7 (-2) & 5.2 \\
Fe-doped & 2.1 (+3) & 3.3 (+2) & 6.9 (-1) & 7.7 (-2) & 6.3 \\
Co-doped & 2.1 (+3) & 3.3 (+2) & 6.8 (-1) & 7.7 (-2) & 7.7 \\ 
Ni-doped & 2.1 (+3) & 3.3 (+2) & 6.8 (-1) & 7.7 (-2) & 9.0 \\
\hline
\end{tabular}
\caption{Resulting Bader analysis after doping the One-Se-plane monolayer.} 
\label{tab:doped-One-Se-plane}
\end{center}
\end{table*}

\end{document}